%%%%%%%% flat currents in the classical        %%%%%%
%%%%%%%% ads_5 x S^5 pure spinor superstring   %%%%%%

\overfullrule=0pt

\input harvmac
\input epsf

%%%%%%%%%%%%%%%% defs %%%%%%%%%%%%%%%%%%

\def\a{{\alpha}}
\def\l{{\lambda}}
\def\b{{\beta}}
\def\g{{\gamma}}
\def\d{{\delta}}

\def\N{{\nabla}}

\def\half{{1\over 2}}
\def\p{{\partial}}
\def\pb{{\bar\partial}}
\def\t{{\theta}}
\def\hat{\widehat}
\def\bar{\overline}

\def\Jb {\bar{J}}
\def\Nb {\bar{N}}
\def\nb {{\bar{\nabla}}}
\def\ah{{\widehat\alpha}}
\def\lh{{\widehat\lambda}}
\def\bh{{\widehat\beta}}
\def\gh{{\widehat\gamma}}

\def\o{{\omega}}
\def\oh{{\widehat\omega}}
\def\th{{\widehat\theta}}

\def\CH{{\cal H}}

\def\CP{{\cal P}}
\def\ll{{\langle}}
\def\rr{{\rangle}}

%%%%%%%%%%%%%%% refs %%%%%%%%%%%%%%%%%%

\lref\berk{N. Berkovits, {\it Super-Poincar\'e
Covariant Quantization of the Superstring,} JHEP 04 (2000) 018,
hep-th/0001035.}

\lref\adswitten{N. Berkovits, C. Vafa and E. Witten,
{\it Conformal Field Theory
of $AdS$ Background with Ramond-Ramond Flux,} JHEP 9903 (1999)
018, hep-th/9902098.}

\lref\bersha{M. Bershadsky, S. Zhukov and A. Vaintrop, {\it $PSL(N | N)$
Sigma Model as a Conformal Field Theory, } Nucl.Phys. B559 (1999) 205,
hep-th/9902180.}

\lref\bz{N. Berkovits, M. Bershadsky, T. Hauer, S. Zhukov and B. Zwiebach,
{\it Superstring Theory on $AdS_2\times S^2$ as a Coset Supermanifold,}
Nucl. Phys. B567 (2000) 61, hep-th/9907200.}

\lref\six{N. Berkovits, {\it Quantization of the
Type II Superstring in a Curved Six-Dimensional Background,}
Nucl. Phys. B565 (2000) 333, hep-th/9908041.}

\lref\metsaev{R. Metsaev and A. Tseytlin, {\it Type
IIB Superstring Action in $AdS_5\times S^5$ Background,}
Nucl. Phys. B533 (1998) 109, hep-th/9805028.}

\lref\chan{N. Berkovits and O. Chand\'{\i}a, {\it Superstring Vertex
Operators in an $AdS_5\times S^5$ Background,} Nucl. Phys. B596 (2001)
185, hep-th/0009168.}

\lref\sugra{N. Berkovits and P. Howe, {\it Ten-Dimensional Supergravity
Constraints from the Pure Spinor Formalism for the Superstring,}
Nucl.Phys. B635 (2002) 75, hep-th/0112160.}

\lref\wave{N. Berkovits, {\it Conformal Field Theory for the
Superstring in a Ramond-Ramond Plane Wave Background,}
JHEP 0204 (2002) 037, hep-th/0203248.}

\lref\adscft{J. Maldacena, {\it The Large N Limit of Superconformal Field
Theories and Supergravity,} Adv.Theor.Math.Phys. 2 (1998) 231,
hep-th/9711200 ; S. S. Gubser, I. R. Klebanov and A. M. Polyakov,
{\it Gauge Theory Correlators from Non-Critical String Theory,}
Phys.Lett. B428 (1998) 105, hep-th/9802109 ; E. Witten,
{\it Anti De Sitter Space And Holography,} Adv.Theor.Math.Phys. 2 (1998) 253,
hep-th/9802150.}

\lref\adsrev{O. Aharony, S. S. Gubser, J. Maldacena, H. Ooguri and Y. Oz,
{\it Large N Field Theories, String Theory and Gravity,}
Phys.Rept. 323 (2000) 183, hep-th/9905111.}

\lref\loren{N. Berkovits and O. Chand\'{\i}a, {\it
Lorentz Invariance of the Pure Spinor BRST Cohomology for the Superstring,}
Phys.Lett. B514 (2001) 394, hep-th/0105149.}

\lref\wadia{Gautam Mandal, Nemani V. Suryanarayana, Spenta R. Wadia,
{\it Aspects of Semiclassical Strings in AdS$_5$,} Phys.Lett. B543 (2002) 81;
hep-th/0206103.}

\lref\oneloop{B. C. Vallilo, {\it One Loop Conformal Invariance of the 
Superstring in an $AdS_5 \times S^5$ Background,} JHEP 0212 (2002) 042, 
hep-th/0210064.}

\lref\abdalla{E. Abdalla, M. C. B. Abdalla and K. D. Rothe, {\it Two 
Dimensional Quantum Field Theory,} World Scientific (Singapore) 2001.}
 
\lref\hidden{I. Bena, J. Polchinski, R. Roiban, {\it Hidden Symmetries 
of the $AdS_5 \times S^5$ Superstring,} hep-th/0305116.}

\lref\work{N. Berkovits, B. C. Vallilo, work in progress.}

%%%%%%%%%%%%%%%% paper %%%%%%%%%%%%%%%%%%%%%%%

\Title{\vbox{\hbox{ IFT-P.029/2003}}}
{\vbox{\centerline{Flat Currents in the Classical $AdS_5 \times S^5$ } 
\smallskip
\centerline{Pure Spinor Superstring }}}
\smallskip
\centerline{Brenno Carlini Vallilo\foot{E-Mail:
vallilo@ift.unesp.br}}
\bigskip
\bigskip
\centerline{\it Instituto de F\'\i sica Te\'orica, Universidade
Estadual Paulista}
\centerline{\it Rua Pamplona 145, 01405-900, S\~ao Paulo, SP, Brasil}

\vskip .3in

It is proven that the classical pure spinor superstring in an 
$AdS_5\times S^5$ background has a flat current depending on a 
continuous parameter. This generalizes the recent result of Bena, 
{\it et al.} for the classical Green-Schwarz superstring.

\Date{July 2003}

%\draft

\newsec{Introduction}

One of the main difficulties of using the quantized superstring in the
$AdS/CFT$ conjecture is the sigma-model describing the dynamics of the 
string in $AdS$ spaces. The RNS superstring is not apropriate to  study 
 RR backgrounds and the covariant quantization of the GS description is still
an open problem. The Berkovits' pure spinor formalism overcomes these two 
difficulties \berk. In the pure spinor description, the  $AdS_5\times S^5$ 
superstring is a coset sigma model plus a WZ term, and in the large radius 
limit quantization can be done without kappa symmetry complications \oneloop. 

Recently, Bena, {\it et al.} \hidden\ have shown that there exist a  
parameter dependent flat current taking values in the $PSU(2,2|4)$ 
algebra in the classical GS superstring description. This flat current 
can be used to construct an infinite number 
of non-local conserved charges. The existence of these charges may be a 
signal that the superstring sigma model might be integrable. Integrability
allows the study of the sigma model outside the large radius limit. In the 
present work, this result will be extended to the pure spinor description 
of the superstring.

There are several questions to be answered before the application of the 
methods of integrable theories to the superstring. The first one is about 
the existence of these flat currents after quantization. Moreover, in general, 
integrable models have a mass gap and asymptotic freedom. This appears to 
be incompatible with the world sheet description, since the sigma model 
must be a conformal field theory and the spatial coordinate is compact. 
The first point is presently under investigation \work.

The result presented here may be easily generalized to the hybrid superstring
descriptions in $AdS_2\times S^2$ \bz\ and $AdS_3\times S^3$ \adswitten\six. 
Note that those descriptions of the superstring are also covariantly 
quantizable. There is also a natural extension of this work to the case of 
pure spinor superstring in a pp-wave background \wave. Since the RR flux in 
that case is not an invertible matrix 
\foot{The flux in the $AdS$ background is proportional to $\gamma^{01234}$, 
$0,1,2,3$ and $4$ being the $AdS$ directions, in the pp-wave, the flux 
is proportional to $\gamma^{-1234}$, where $-$ is the light-cone direction.}, 
the world sheet fields $(d_\a,\hat d_\ah)$ cannot be eliminated by their 
equations of motion and they must be included in the flat current.   

This paper is organized as follows. In section 2 a short review of the pure
spinor superstring is given. The construction of the flat current is done 
in section 3. The Appendices have short derivations 
that were omitted in the body of the paper.

\newsec{Pure Spinor Superstring in an $AdS_5\times S^5$ Background}

In this section I review the pure spinor formalism for the  
$AdS_5 \times S^5$ background as a coset supermanifold.

As was shown in \metsaev\ , the  $AdS_5 \times S^5$ background can be
described by a coset supergroup element $g$ taking values in
$PSU(2,2|4)/SO(4,1)\times SO(5)$ where the supervierbein and spin
connections are given by

$$E^A_M dy^M = (g^{-1}dg)^A = J^A, $$
where $A=(\underline a,\a,\ah,[\underline a \underline b])$ and
$\underline{a}$ signifies either $a$ or $a'$ and
$[\underline{cd}]$ signifies either $[ab]$ or $[a'b']$, $a=0$ to $4$
and $a'=5$ to $9$. The non-vanishing structure constants $f_{AB}^C$
of the $PSU(2,2|4)$ algebra are
\eqn\structure{
f_{\a\b}^{\underline{c}} =2 \g^{\underline{c}}_{\a\b},\quad
f_{\ah\bh}^{\underline{c}} =2 \g^{\underline{c}}_{\a\b},}
$$
f_{\a \bh}^{[ef]}=
f_{\bh \a}^{[ef]}=
(\g^{ef})_\a{}^\g \d_{\g\bh},\quad
f_{\a \bh}^{[e'f']}=
f_{\bh \a}^{[e'f']}= -
(\g^{e'f'})_\a{}^\g \d_{\g\bh},$$
$$f_{\a \underline{c}}^\bh
=-f_{\underline{c}\a}^\bh
=\half (\g_{\underline c})_{\a\b}
\d^{\b\bh},\quad
f_{\ah \underline{c}}^\b =
-f_{\underline{c}\ah}^\b =
-\half
(\g_{\underline c})_{\ah\bh} \d^{\b\bh},$$
$$f_{c d}^{[ef]}= \half \d_c^{[e} \d_d^{f]},
\quad f_{c' d'}^{[e'f']}= -\half \d_{c'}^{[e'} \d_{d'}^{f']},$$
$$f_{[\underline{cd}][\underline{ef}]}^{[\underline{gh}]}=\half (
\eta_{\underline{ce}}\d_{\underline{d}}^{[\underline{g}}
\d_{\underline{f}}^{\underline{h}]}
-\eta_{\underline{cf}}\d_{\underline{d}}^{[\underline{g}}
\d_{\underline{e}}^{\underline{h}]}
+\eta_{\underline{df}}\d_{\underline{c}}^{[\underline{g}}
\d_{\underline{e}}^{\underline{h}]}
-\eta_{\underline{de}}\d_{\underline{c}}^{[\underline{g}}
\d_{\underline{f}}^{\underline{h}]})$$
$$f_{[\underline{cd}] \underline{e}}^{\underline{f}} =
-f_{\underline{e} [\underline{cd}]}^{\underline{f}}= \eta_{\underline{e}
\underline{[c}} \d_{\underline d]}^{\underline{f}},\quad
f_{[\underline{cd}] \a}^{\b} =
-f_{\a [\underline{cd}]}^{\b}= \half(\g_{\underline{cd}})_\a{}^\b,\quad
f_{[\underline{cd}] \ah}^{\bh} =
-f_{\ah [\underline{cd}]}^{\bh}= \half(\g_{\underline{cd}})_\ah{}^\bh.$$

The $PSU(2,2|4)$ algebra $\CH$ has a natural decomposition \bz\
$\CH=\sum \CH_i+ \CH_0$, $i=1$ to $3$ \eqn\dec{J_{\underline a} \in
\CH_2, \quad J_{[\underline{ab}]} \in \CH_0, \quad J_\a \in \CH_1,
\quad J_\ah \in \CH_3.}

As it can be seen from the structure constants \structure\ 
\eqn\alge{[\CH_0,\CH_0]\subset \CH_0,\quad [\CH_0,\CH_i]\subset \CH_i,
\quad [\CH_i ,\CH_j ] \subset \CH_{i+j} \quad {\rm mod }~4.} The
bilinear form also respects the decomposition 
\eqn\bili{ \ll \CH_0,\CH_i \rr =0,\quad \ll \CH_0,\CH_0\rr \neq 0, 
\quad \ll \CH_i , \CH_j \rr =0\quad {\rm unless }\quad i+j=0 
\; ({\rm mod }~4).}

The action for the pure spinor superstring in this background is 
\berk\chan\oneloop 

\eqn\currentact{S_{AdS}= \int d^2 z [
 \ll J_2 ,\bar J_2 \rr + {3\over 2}\ll J_3 , \bar J_1 \rr +
{1\over 2}\ll\bar J_3,J_1 \rr ] + }
$$ + \int d^2 z [ N_{\underline c \underline d}
\overline J^{[\underline c \underline d]} +
 \Nb_{\underline c \underline d} J^{[\underline c \underline d]}
+ \half N_{\underline c \underline d} \Nb^{\underline c \underline d} ]
+ S_\l + S_{\widehat\l}, $$
where $J^A = (g^{-1} \p g)^A$ and $\overline J^A = (g^{-1} \overline\p
g)^A$ are left-invariant currents constructed from the supergroup element
$g \in PSU(2,2|4)$ and $J_i=J\big|_{\CH_i}$.  
$S_\l$ and $S_\lh$ are the free field actions
for the bosonic left and right-moving ghosts $\l^\a$ and $\lh^\ah$
satisfying the pure spinor conditions

\eqn\pure{ \l\g^{\underline a}\l=0 {\rm ~~and ~~} \lh\g^{\underline a}\lh=0.}

Although an explicit form  of $S_\l$ and $S_\lh$ exist, it will not be 
needed here. $(N^{cd},N^{c'd'})$ and
$(\Nb^{cd},\Nb^{c'd'})$ are the $SO(4,1)\times SO(5)$ 
components of the Lorentz current for the bosonic ghosts.
The OPE's of $\l^\a$ and $\lh^\ah$ with these
currents are manifestly $SO(4,1)\times SO(5)$  covariant. Instead of having 
kappa symmetry, the action is BRST invariant with charges given by
$Q=\oint \l^\a J_1^\ah \d_{\a\ah}$ and 
$\bar Q=\oint \lh^\ah \Jb_3^\a \d_{\a\ah}$ \chan. Physical states are vertx 
operators in the cohomology of $Q$ and $\bar Q$.

Under a local $SO(4,1)\times SO(5)$ transformation parametrized by 
$\Omega \in \CH_0$, 
\eqn\local{\delta J_i = [J_i,\Omega],\quad \delta \Jb_i = [\Jb_i,\Omega],
\quad \delta J_0 = \p \Omega + [J_0,\Omega], \quad 
\delta \Jb_0 = \pb \Omega + [\Jb_0,\Omega],}
$$\delta (S_\l + S_\lh)= -\p \Omega^{\underline a \underline b} 
N_{\underline a \underline b} - \pb \Omega^{\underline a \underline b} 
\Nb_{\underline a \underline b}.$$
$$ \delta N = [N,\Omega],\quad \delta \Nb = [\Nb,\Omega].$$

To calculate the flatness condition one must know the equations of 
motion that follows from \currentact. Under an arbitrary variation 
$\delta g= gX$, $\delta g^{-1}=-Xg^{-1}$ with $X \in \CH_i$,
\eqn\vari{\delta J = \p X + [J,X] , \quad 
\delta \Jb =\pb X + [\Jb,X] .}
Together with the Maurer-Cartan equations $\p \Jb -\pb J + [J,\Jb]=0$, 
\vari\ implies the equations of motion
\eqn\eqmov{\N\Jb_2= - [J_3,\Jb_3] -\half [N,\Jb_2]+\half [J_2,\Nb],}
$$\nb J_2 = [J_1,\Jb_1] +\half [J_2,\Nb]-\half [N,\Jb_2],$$
$$\nb J_1 = \half [N,\Jb_1]-\half [J_1,\Nb],$$
$$\N \Jb_1 = -[J_2,\Jb_3]-[J_3,\Jb_2]+\half [N,\Jb_1]-\half [J_1,\Nb],$$
$$\N \Jb_3 = \half [N,\Jb_3]-\half [J_3,\Nb],$$
$$\nb J_3 = [J_2,\Jb_1]+[J_1,\Jb_2]+\half [N,\Jb_3]-\half [J_3,\Nb],$$
where $\N=\p + [J_0, \;\; ]$ and $\nb=\pb + [\Jb_0, \;\; ]$.

Instead of using $(\l,\lh)$ and their conjugate momenta $(\o,\oh)$ to derive
the equation of motion for $(N,\Nb)$, this can be done using the gauge 
invariance \local\ ignoring the variation of $(J_0,\Jb_0)$. This is because
\local\ is the most general $SO(4,1)\times SO(5)$ covariant variantion of 
$N$ and $\Nb$. The gauge symmetry is a special case. 
Varying the second line of \currentact\ with independent $\Lambda$ and 
$\bar \Lambda$ gives 
\eqn\eqmovn{\nb N = \half [N,\Nb],\quad \N \Nb = -\half [N,\Nb].}

The pure $AdS_5 \times S^5$ spinor superstring was first presented in the
paper \berk. In \chan, the massless vertex operator corresponding to
supergravity fluctuations around the $AdS$ background was studied. It was 
shown that the classical BRST currents $\l^\a J^\ah_1 \d_{\a\ah}$ and 
$\lh^\ah \Jb^\a_3 \d_{\a\ah}$ are holomorphic and antiholomorphic and their
corresponding BRST operators anticommute. One loop conformal invariance 
of \currentact\ was proved in \oneloop.

\newsec{Flat Currents in the Classical Pure Spinor Superstring}

In a recent paper, Bena {\it et al.} \hidden\ showed that the classical
GS superstring action in the $AdS_5\times S^5$ background has a one 
parameter dependent flat current. This flat current can be used to construct
infinitely many conserved charges. The existence of these charges is an 
indication that the model might be integrable. Although the GS superstring
is a well defined classical theory, its quantization proves to be 
a major problem and it is only well understood in the lighcone gauge.

In this section I will extend the results of \hidden\ to the case of the
pure spinor superstring. Under the the $\CH_0$-gauge transformation 
$g\to gh$, $J_0$ and $\bar J_0$ 
transform as connections and $J_i$, $N$ and $\Jb_i$, $\Nb$ 
transform in the adjoint $\CH_0$-representation. This means that the currents
$g J_i g^{-1}$, $g N g^{-1}$ and $g \Jb_i g^{-1}$, $g \Nb g^{-1}$
are invariant under the $\CH_0$ transformation.

If it is possible to find a $\mu$ dependent $\CH_0$ invariant current 
$a(\mu)$ that satisfies the flatness condition
\eqn\flat{da(\mu) + a(\mu) \wedge a(\mu) = 0,}
an infinit number of conserved charges can be constructed. 
See the Appendix 2 for a short derivation.

Since the left invariant currents $J_i$ and $\Jb_i$ are easier to handle, 
it is useful to write \flat\ in terms of a left invariant 
$A(\mu)=g^{-1} a(\mu) g$. The last term in \flat\ is covariant under this 
transformation, but $d a(\mu)$ is transformed to
$$d(g^{-1}a(\mu)g)= -g^{-1}dg\wedge(g^{-1}a(\mu)g) +g^{-1}da(\mu)g - 
(g^{-1}a(\mu)g)\wedge g^{-1}dg.$$
Writing in terms of $A$ and $J=\sum J_i + J_0$, 
$$g^{-1}da(\mu)g = dA(\mu)+ J\wedge A(\mu) + A(\mu)\wedge J.$$ 
\flat\ is now writen as
\eqn\aflat{ dA(\mu) + J\wedge A(\mu) + A(\mu)\wedge J + 
A(\mu)\wedge A(\mu) = 0.}
In the $z,\bar z$ notation,
\eqn\flat{\p \bar A - \pb A + [A,\bar A] + [J,\bar A] + [A, \bar J]=0.}

One aspect that makes the pure spinor superstring quantizable is that there 
are equations of motion for all the $J_i$ and $\bar J_i$. This 
enables us to construct a flat connection $A$ in terms of all of them 
$$A= a J_2 + bJ_1 + cJ_3,$$
$$\bar A = d \Jb_2 + e\Jb_1+f\Jb_3.$$
In \hidden\ only antisymmetric combinations of $(J_1,\Jb_1)$ and 
$(J_3,\Jb_3)$ could be included in $A(\mu)$. In the GS superstring kappa 
symmetry is an essential ingredient and restricts the form of the action.  
The equations of motion for $(J_1,\Jb_1)$ and $(J_3,\Jb_3)$ 
in the the GS model are
$$[J_2,\Jb_3]=-[J_3,\Jb_2],\quad [J_2,\Jb_1]=-[J_1,\Jb_2],$$
with no terms containing covariant derivatives. Only the Maurer-Cartan 
identity can be used in \flat\ before fixing kappa symmetry. 

The quantizable action also has the bosonic pure spinor ghosts. These 
ghosts have Lorentz currents $N$ and $\Nb$ that interact 
with the sigma model, so they have to be included in $A$ and $\bar A$
\eqn\aandabar{A= a J_2 + bJ_1 + cJ_3 + gN,}
$$\bar A = d \Jb_2 + e\Jb_1+f\Jb_3 + h\Nb.$$

Now \flat\ is expanded using $A$ and $\bar A$ given above.
\eqn\equation{ d\N \Jb_2 + e\N\Jb_1 + f\N\Jb_3 +h\N\Nb - 
a\nb J_2 -b\nb J_1 -c\nb J_3 -g\nb N +}
$$+ (ad+d+a)[J_2,\Jb_2] + (ae+e+a)[J_2,\Jb_1] + (af+f+a)[J_2,\Jb_3] +
(bd+d+b)[J_1,\Jb_2] + $$
$$+(be+b+e)[J_1,\Jb_1] + (bf+b+f)[J_1,\Jb_3] + (cd+c+d)[J_3,\Jb_2] +
(ce+c+e)[J_3,\Jb_1] + $$
$$ + (cf+c+f)[J_3,\Jb_3]+(ha+h)[J_2,\Nb]+(hb+h)[J_1,\Nb]+(hc+h)[J_3,\Nb] +$$
$$+ (gd+g)[N,\Jb_2]+(ge+g)[N,\Jb_1]+(gf+g)[N,\Jb_3]+gh[N,\Nb]=0,$$

Inserting the equations of motion \eqmov\ in \equation, the following 
system of equations involving all coefficients $(a,b,c,d,e,f,g,h)$ 
has to be solved

\eqn\equations{ad+d+a=0,\quad ea+e+a-c=0,\quad af+f+a-e=0,}
$$ bd+d+b-c=0,\quad be+b+e-a=0,\quad bf+b+f=0,$$
$$ cd+d+c-e=0,\quad ce+c+e=0,\quad cf+c+f-d=0,$$
$$ gh-\half(g+h)=0,\quad ha+h-\half(a-d)=0,\quad hb+h-\half(b-e),$$
$$ hc+h-\half(c-f)=0,\quad gd+g-\half(d-a)=0,\quad ge+g-\half(e-b)=0,$$
$$gf+g-\half(f-c)=0.$$

The solution is given by
\eqn\solution{a=\mu-1,\quad b=\pm(\mu)^{3\over 2}-1,\quad 
c=\pm(\mu)^{\half}-1,}
$$d=\mu^{-1}-1,\quad e=\pm(\mu)^{-\half}-1,
\quad f=\pm(\mu)^{-{3\over 2}}-1,$$
$$g=\half(1-\mu^2),\quad h=\half(\mu^{-2}-1),$$
which is analytic for $\mu >0$. It is interesting to note that the system 
has the same solution if we exclude the ghost currents. This is not 
surprising, since the ghosts are necessary to quantization and the 
calculation here is classical. Note that this independence means that exactly 
the same result holds for the hybrid superstring description in 
$AdS_2\times S^2$ and $AdS_3\times S^3$. When quantum corrections are 
taken into account, the ghost part has an essential role \work.

To obtain the result of Bena {\it et al.} it must be remembered that
the pure spinor superstring action \currentact\ can be writen in the form
$$S_{PS}= S_{GS}+ \int d^2z(d_1 \Jb_3 +\bar d_3 J_1 + d_1\bar d_3) 
+S_{ghosts}$$
with the indexes contracted in a $PSU(4|4)$ covariant way. The GS superstring
is recovered with $d_1=0$ and $\bar d_3=0$ and $S_{ghosts}=0$. 
Calculating the equations of motion, this is equivalent 
to $\Jb_3=0$ and $J_1=0$. Doing that, we end up with the equations
\eqn\bena{ad+d+a=0,\quad ae+a+e-c=0, \quad cd+c+d-e=0, \quad ce+c+e=0.}

Writing $(a,c,d,e)$ in terms of $(\a,\b,\g,\d)$,
$$a=\b-\a,\quad c=\d-\g,\quad d=-\b-\a,\quad e=-\g-\d,$$
\bena\ is
$$\a^2-\b^2-2\a=0,\quad -\b\g+\a\g-\b\d+\a\g+\b-\a-2\d=0$$
$$\g^2-\d^2-2\g=0,\quad -\b\d+\b\g+\g\a-\d\a-\a-\b+2\d=0 ,$$
which are linear combinations of their main equations.

\vskip 15pt {\bf Acknowledgements:} I would like to thank L. A. Ferreira
for useful discussions and especially Nathan Berkovits for useful discussions
and suggestions. I also thank D\'afni Marchioro for a careful reading of the 
manuscript. This work was supported by FAPESP grant 00/02230-3.

\newsec{Appendix 1}

The action \currentact\ is derived from a general curved background

\eqn\curvact{S_{curv}= \int d^2 z [ \half(G_{MN}+B_{MN})\p y^M \bar\p y^N + }
$$ + d_\a E^\a_M\bar \p y^M + \half N_{mn}\pb y^M \Omega_M^{mn} +
\hat d_\ah \hat E^\ah_M \p y^M +\half \bar N_{mn} \p y^M \hat \Omega_M^{mn} +
d_\a \hat d_\gh P^{\a\gh} + $$
$$+ \bar N_{mn}d_\a C^{\a mn} +
 N_{mn}\hat d_\ah \hat C^{\ah mn}+ {1\over 2}N_{mn}\bar
N_{op}R^{mnop} + \Phi(x,\t,\th)R \quad ] + S_{\l} + S_{\lh},$$
where the superfields $(G_{MN},B_{MN},E^\a_M,\hat
E^\ah_M,\Omega_M^{mn},\hat\Omega_M^{mn}, P^{\a\gh},C^{\a mn},\hat
C^{\ah mn},R^{mnop},\Phi)$ are related to the supergravity
multiplet \sugra. The first line in \curvact\ is just the GS
action in a curved background, the second and third lines are
necessary to covariantly quantize the superstring, {\it i.e.}, 
they provide invertible propagators (containing no operators that
might have zero modes) for the fermions. In flat space, the world sheet 
fields $(d_\a,\hat d_\ah)$ are expressed in terms of $(x,\t,\th)$, but in 
a general background they are interpreted as independent fields. 
$\Phi(x,\t,\th)R$ is the
Fradkin-Tseytlin term, where $\Phi$ is the compensator scalar
superfield whose lowest component is the dilaton and $R$ is the
worldsheet curvature. Since the dilaton is constant in the 
$AdS_5 \times S^5$ background, the Fradkin-Tseytlin term is integrated 
to give the usual genus counting coupling constant.

To get \currentact, one must know the value of the background superfields
$B_{AB}$ and $P^{\a\bh}$
\eqn\back{B_{\a\bh}=B_{\bh\a}=-\half (Ng_s)^{1\over 4}\d_{\a\bh},
\quad P^{\a\bh}={1\over (Ng_s)^{1\over 4}} \d^{\a\bh},} where $N$
is the value of the Ramond-Ramond flux, $g_s$ is the string
coupling constant and $\d_{\a\bh}=(\g^{01234})_{\a\bh}$ with 01234
being the directions of $AdS_5$. The background field $P^{\a\bh}$ makes
$(d_\a,\hat d_\ah)$ auxiliary fields which can be eliminated by their
equations of motion.

\newsec{Appendix 2}

In this Appendix, a short derivation of the conserved charge implied 
by the flatness condition is given.\foot{See, 
for example, \abdalla\ for a review}

It folows by the Non-abelian Stokes Theorem that the path ordered integral is
\eqn\stokes{U_\omega [a]=\CP e^{\oint_\omega a(\mu)}=1,}
for any contractible loop $\omega$. In radial quantization, time 
evolution is represented by the radial coordinate. The origin is the 
infinite past, where usually there is some vertex operator inserted. Because 
of this operator insertion, loops around the origin are not contractible. 
The path ordered integral $\CP e^{\oint_\rho a(\mu)}$ satisfies the 
group product rule, 
$$U_\rho [a] U_\sigma [a] =U_{\sigma\circ\rho} [a],$$
if the end of $\sigma$ is the begginning of $\rho$.

\vskip 15pt
\centerline{\epsfbox{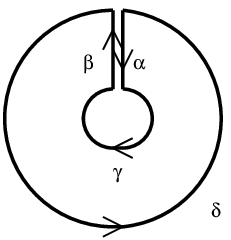}}
\centerline{Figure 1: Path chosed to prove conservation of }
\centerline{$Tr(U_\g [a])$ for non-contractible loops.}
\vskip 15pt

Choosing the path in Figure 1, 
$$U_\a [a] U_\d [a] U_\b [a] U_\g [a]= U_{\g\circ\b\circ\d\circ\a} [a]=1.$$
From this equation follows that 
\eqn\conser{Tr(U_\g [a])= Tr(U_{\d^{-1}} [a]),}
which means that $Tr(U_\g [a])$ is a conserved charge. Since the flat 
current $a$ depends on a continuous parameter, there are an infinite number 
of conserved charges.

\listrefs
\end